\documentstyle[12pt]{article}

\def \beq { \begin{equation} }
\def \eeq { \end{equation} }

\begin{document}
\input{epsf.tex}

\thispagestyle{empty}

\begin{center}
{\Large \bf Oscar's Physics Phaire} \\
{\bf A Collection of Problems}
\end{center}
\vskip .7 cm
\noindent
{\it You will find here a number of (mostly) elementary physics 
problems dealing mainly with uniform motion kinematics. In 
preparing this collection I have tried to create original 
situations that could help bring motivation to an introductory
course. You are welcome to suggest improvements and ... 
{\sl provide  solutions!} 
\begin{flushright}
{\small Oscar Bolina\\
Department of Mathematics \\
University of California, Davis\\
Davis, CA 95616-8633\\
{\bf E-mail:} bolina@math.ucdavis.edu}\\
\end{flushright}
}

%%%%%%%%%%%%%%%%%%%%%%%%%%%%%%%%%%%%%%%%%%%%%%%%%%%%%%%%

\vskip 2 cm
\begin{center}
{\large The Four Spiders}
\end{center}
\noindent
\vskip .2 cm
\noindent
Four spiders move in straight lines away from a 
common origin on a plane in such a way that at 
any time they are situated on the corners of a 
rectangle. If three spiders move at rates of 2, 
3 and 4 cm/s, find out the possible values of 
the speed of the fourth spider.

%%%%%%%%%%%%%%%%%%%%%%%%%%%%%%%%%%%%%%%%%%%%%%%%%%%%%%%%%%%%%

\vskip 2 cm
\begin{center}
{\large Dangerous Nanny}
\end{center}
\noindent
\vskip .2 cm
\noindent
On his everyday commute from home to work 
at 50 km/h, a man always met his nanny, 
heading to his house, halfway on his
journey to work. One day the man left home 
5 minutes later and had to drive at 70 km/h 
in order to get to work on time. On this day, 
he met his nanny 9 km from his house. What 
is the speed the nanny used to maintain on 
her daily ride?

%%%%%%%%%%%%%%%%%%%%%%%%%%%%%%%%%%%%%%%%%%%%%%%%%%%%%%%%%%%%%%

\vskip 2 cm
\begin{center}
{\large A Sliding Problem}
\end{center}
\noindent
\vskip .2 cm
\noindent
A bead slides with constant speed {\it v} along 
lines perpendicular to the sides of a cone of 
side {\it l} and base {\it b}. How long does it 
take for the bead to reach the vertex {\it V}?
\begin{figure}
\centerline{
\epsfbox{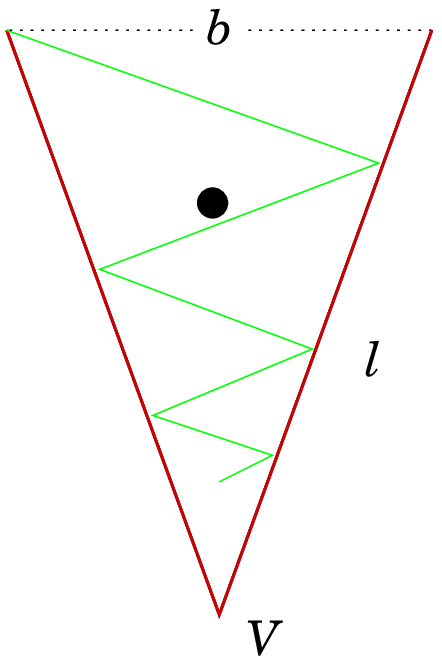}}
%\caption{}
\end{figure}

%%%%%%%%%%%%%%%%%%%%%%%%%%%%%%%%%%%%%%%%%%%%%%%%%%%%%%%%%%

\vskip 2 cm
\begin{center}
{\large Cubic River}
\end{center}
\noindent
\vskip .2 cm
\noindent
A boat that can move at speed {\it v} in still water crosses a 
river of width {\it l} that flows with speed {\it w}. Prove that 
the shortest time the boat takes to complete its trip is $l/v$.
Show that the boat reaches the opposite bank at a distance 
$x=wl/v$ {\it (above or below?)} from its starting position.
Verify that this value for {\it x} is a root of the equation
\[
v^{2} x^{6} + l^{2}(2v^{2}-w^{2})x^{4}+l^{4}(v^{2}-2w^{2})x^{2} 
-w^{2}l^{6}=0.
\]
Solve this (cubic) equation for other roots and interpret your result.

%%%%%%%%%%%%%%%%%%%%%%%%%%%%%%%%%%%%%%%%%%%%%%%%%%%%%%%%%%%%%%%%%%%%

\vskip 2 cm
\begin{center}
{\large Bugs' Lives}
\end{center}
\noindent
\vskip .2 cm
\noindent
A bug {\it A} that was confined to live on the rim of a circle of 
radius {\it R} realized that a tasty bug {\it B} it preyed on used 
to sneak into its territory always with the same constant velocity 
{\it v} along a same straight line located at a distance {\it r} 
from the enter of the circle. Bug {\it A} would like to figure out 
the minimum velocity it should start moving as soon as bug {\it B} 
showed up at {\it p} so that, no matter where it was on the 
circumference of the circle, it would always arrive at {\it q} before 
bug {\it B}, and, hopefully, have a good snack. Could you help?
\begin{figure}
\centerline{
\epsfbox{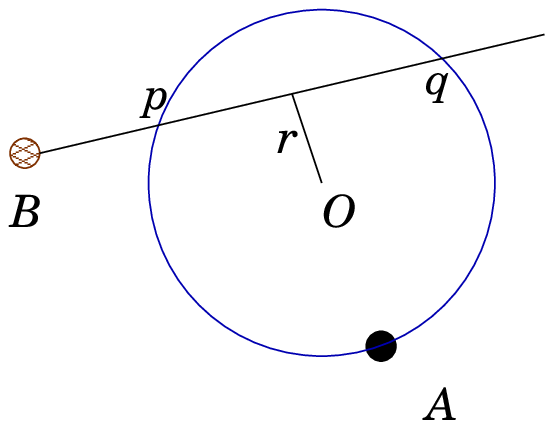}}
%\caption{}
\end{figure}

%%%%%%%%%%%%%%%%%%%%%%%%%%%%%%%%%%%%%%%%%%%%%%%%%%%%%%%%%%

\vskip 2 cm
\begin{center}
{\large The Closer the Faster}
\end{center}
\noindent
\vskip .2 cm
\noindent
Two particles describe a rectilinear motion in a 
plane in such a way that the component of their 
relative velocity along the straight line joining 
them has a constant magnitude {\it u}. Prove that 
their relative velocity as a function of the 
distance {\it d} between them is given by 
\beq
V=\frac{u}{\sqrt{1-({d_{min}}/d)^{2}}}
\eeq
where $d_{min}$ is the minimum distance between 
the particles

%%%%%%%%%%%%%%%%%%%%%%%%%%%%%%%%%%%%%%%%%%%%%%%%%%%%%%%%%%%%%
\newpage
\vskip 2 cm
\begin{center}
{\large Square Dance}
\end{center}
\noindent
\vskip .2 cm
\noindent
Four ants, initially located at the corners of a square of side 
{\it d}, start to move at the same time, in the same sense and 
with constant speeds, along the sides of the square. How long 
does it take for the ants to be, for the first time, all moving 
on the same side of the square? Are there conditions on the 
speeds for this to be possible? 
\begin{figure}
\centerline{
\epsfbox{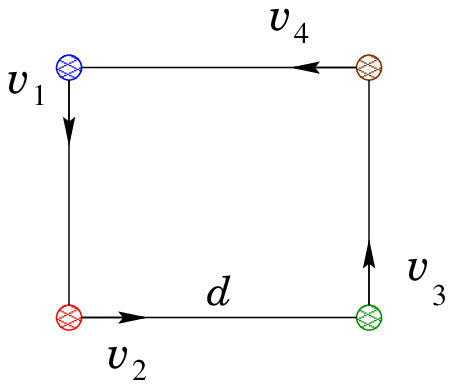}}
%\caption{}
\end{figure}

%%%%%%%%%%%%%%%%%%%%%%%%%%%%%%%%%%%%%%%%%%%%%%%%%%%%%%%%%%%%
 
\vskip 2 cm
\begin{center}
{\large Probabilistic Kinematics}
\end{center}
\noindent   
\vskip .2 cm
\noindent   
Particles are created at random on a unit line {\it AB}
and move towards {\it B} with speed {\it v}. What is the 
probable elapsed time between creation and detection at 
{\it B}?

%%%%%%%%%%%%%%%%%%%%%%%%%%%%%%%%%%%%%%%%%%%%%%%%%%%%%%%%%%%%%%%%

\vskip 2 cm
\begin{center}
{\large Storm Ahead}
\end{center}
\noindent
\vskip .2 cm
\noindent
Forty minutes into its straight line flight of 1,800 miles to
Clear Skies City at 360 miles an hour, an airplane is warned 
against a storm ahead and ordered to take an alternative route
and increase its speed $30 \%$. Determine how far from its routine 
route can the plane get without delaying itself.

%%%%%%%%%%%%%%%%%%%%%%%%%%%%%%%%%%%%%%%%%%%%%%%%%%%%%%%%%%%%%%%

\vskip 2 cm
\begin{center}
{\large Of Life and Death}
\end{center}
\noindent
\vskip .2 cm
\noindent
A photon, created at {\it P} with speed {\it c}, 
reflects twice on the walls of a square box before
being absorbed at the origin {\it O}. How long does
the photon survive?
\begin{figure}
\centerline{
\epsfbox{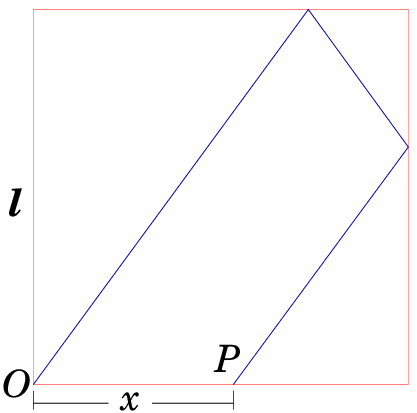}}
%\caption{}
\end{figure}

%%%%%%%%%%%%%%%%%%%%%%%%%%%%%%%%%%%%%%%%%%%%%%%%%%%%%%%%%%%%%

\vskip 2 cm
\begin{center}
{\large Squirrels Are Not Pets}
\end{center}
\noindent
\vskip .2 cm
\noindent
A very fearful squirrel is careful enough never to go farther 
than a distance {\it R} from its burrow. One day the squirrel 
leaves its burrow, gets a cone near a pine tree, gets a nut
near an oak tree and returns to the burrow leaving behind 
a path in the shape of a triangle having maximum area. If the 
squirrel can run with speed {\it v}, how much time does it spend 
on the round-trip?

%%%%%%%%%%%%%%%%%%%%%%%%%%%%%%%%%%%%%%%%%%%%%%%%%%%%%%%%%%%%%%%

\vskip 2 cm
\begin{center}
{\large Fast and First}
\end{center}
\noindent
\vskip .2 cm
\noindent
Two particles, initially at a distance {\it r} apart, move for 
a time {\it T} with speeds {\it v} and {\it w} along straight 
lines towards the same point. Determine all the points that the 
faster particle reaches before the slower one.
 
%%%%%%%%%%%%%%%%%%%%%%%%%%%%%%%%%%%%%%%%%%%%%%%%%%%%%%%%%%%%%%%

\vskip 2 cm
\begin{center}
{\large Askance and ...}
\end{center}
\noindent
\vskip .2 cm
\noindent
Two particles that move in a plane along intersecting 
straight lines with constant speeds $v$ and $w$ have 
the component of these velocities along the line 
joining the particles always in the ratio 1:k. Prove 
that angle $\beta$ between the directions of motion 
of the two particles is given by  
\[
\cos \beta= \frac{v^{2}-kw^{2}}{(1-k)vw}
\]

%%%%%%%%%%%%%%%%%%%%%%%%%%%%%%%%%%%%%%%%%%%%%%%%%%%%%%%%%%%%%%%%

\vskip 2 cm
\begin{center}
{\large ... Oblique}
\end{center}
\noindent   
\vskip .2 cm
\noindent   
The components of the acceleration of a particle along two direction
in the plane of the acceleration vector that make an angle $\xi$ with 
each other are $a_{1}=ap$ and $a_{2}=aq$, where $a$ is the magnitude 
of the acceleration vector. Show that 
\[
\sin{\xi}=\frac{1}{2pq} \sqrt{2(p^{2}+q^{2})-(p^{2}-q^{2})^{2}-1}.
\]
Are there restrictions on the values of $p,q$?

%%%%%%%%%%%%%%%%%%%%%%%%%%%%%%%%%%%%%%%%%%%%%%%%%%%%%%%%%%%%%%%

\vskip 2 cm
\begin{center}
{\large Under Arrest}
\end{center}
\noindent
\vskip .2 cm
\noindent
When the police arrived at a bank responding to a robbery,
the robbers had already fled in two cars speeding at 80 km/h
and 90 km/h. The police pursue the cars, arresting the slower 
car at 20 km from the bank and the faster car at 30 km from the 
bank. Determine the speed of the police car in the pursuit.

%%%%%%%%%%%%%%%%%%%%%%%%%%%%%%%%%%%%%%%%%%%%%%%%%%%%%%%%%%%%%%%

\newpage
\vskip 2 cm
\begin{center}
{\large Math Illusion}
\end{center}
\noindent
\vskip .2 cm
\noindent
A particle moves with speed {\it v} on
concentric triangular lines having lengths 
$p \%$ shorter than the previous one. If
the longest line has unit length, how long 
does it take for the particle to reach the 
center?
\begin{figure}
\centerline{
\epsfbox{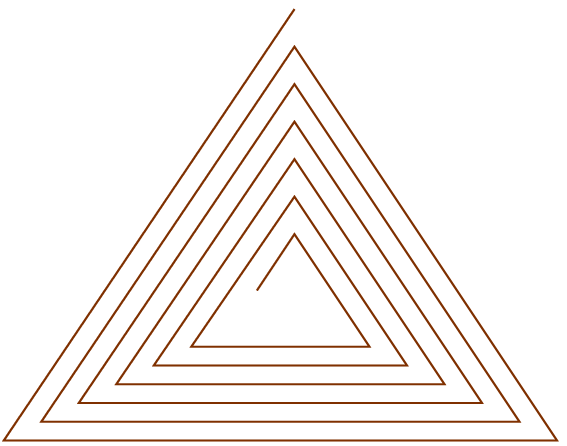}}
%\caption{}
\end{figure}
 
%%%%%%%%%%%%%%%%%%%%%%%%%%%%%%%%%%%%%%%%%%%%%%%%%%%%%%%%%%%%%%%%

\vskip 2 cm
\begin{center}
{\large Downstreaming}
\end{center}
\noindent   
\vskip .2 cm
\noindent   
A boat capable of developing speed {\it v} in still water
crosses a river of width {\it l} and reaches the opposite
shore at a distance {\it x} below its starting position. 
Determine the speed of the current that minimizes the time
of travel.

%%%%%%%%%%%%%%%%%%%%%%%%%%%%%%%%%%%%%%%%%%%%%%%%%%%%%%%%%%%%%%%%%%

\vskip 2 cm
\begin{center}
{\large Here and There}
\end{center}
\noindent
\vskip .2 cm
\noindent
A particle moves in the plane defined by two straight lines 
that meet at an angle $\psi$. If the distances of the particle 
to both lines are always in the ratio $1:r$, show that the 
components of its velocity along the two lines are in the 
ratio 
\[
\frac{r+\cos{\xi}}{1+r\cos{\xi}}
\]

%%%%%%%%%%%%%%%%%%%%%%%%%%%%%%%%%%%%%%%%%%%%%%%%%%%%%%%%%%%%%%%%

\vskip 2 cm
\begin{center}
{\large Work Out}
\end{center}
\noindent
\vskip .2 cm
\noindent
You are willing to burn some calories {\it (without much effort)} 
and decide to run from some point {\it A} to {\it B}, {\it C} and 
back to {\it A}, all on the side of an equilateral triangle of 
length {\it l}. Before starting, find out which initial position 
{\it x} minimizes your route.
\begin{figure}
\centerline{
\epsfbox{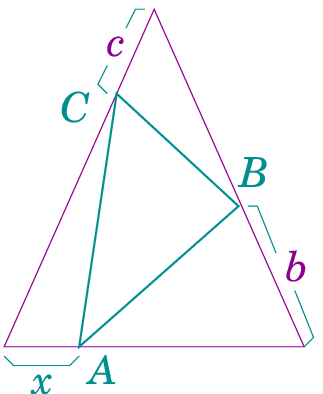}}
%\caption{}
\end{figure}

%%%%%%%%%%%%%%%%%%%%%%%%%%%%%%%%%%%%%%%%%%%%%%%%%%%%%%%%%%%

\noindent
\vskip 2 cm
\begin{center}
{\large A Train Breakdown}
\end{center}
\noindent
\vskip .2 cm
\noindent
The trains that serve two stations 60 km apart leave at regular
intervals of 20 min. One day, one of the trains experiences 
mechanical problems after traveling 40 km and has to complete
the rest of the trip at half its usual speed. As a consequence,
this train arrives at the other station just 8 min before the next
train. Determine the usual speed of the trains.

%%%%%%%%%%%%%%%%%%%%%%%%%%%%%%%%%%%%%%%%%%%%%%%%%%%%%%%%%%%%%%%%%%%%%

\vskip 2 cm
\begin{center}
{\large Animal Procession}
\end{center}
\noindent
\vskip .2 cm
\noindent
A hen, a pig, and a dog leave a barn at equal time 
intervals and with speeds $p, q, r$ and arrive at 
the farmhouse also at equal time intervals but in 
inverse order. Show that 
\[
\frac{1}{p} + \frac{1}{r}=\frac{2}{q}
\]

%%%%%%%%%%%%%%%%%%%%%%%%%%%%%%%%%%%%%%%%%%%%%%%%%%%%%%%%%%%%%%%%%

\vskip 2 cm
\begin{center}
{\large Bouncing Molecule}
\end{center}
\noindent
\vskip .2 cm
\noindent
An air molecule on the corner of a container makes its way out 
of the container by progressively bouncing off its walls until 
it escapes through the opening. Determine the initial directions 
of the molecule's velocity vector that allow an escape in the
case the molecule experiences no loss of speed. How would you 
change your answer if the molecule started off with speed $v_{0}$ 
and lost one third of it after each strike on the wall? What is
the minimum possible value of $v_{0}$ in this case?
\begin{figure}
\centerline{
\epsfbox{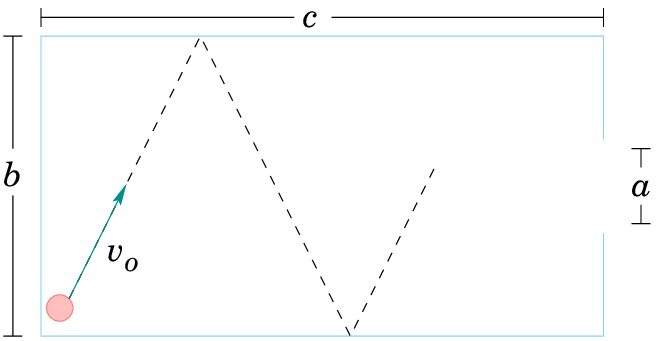}}
%\caption{}
\end{figure}

%%%%%%%%%%%%%%%%%%%%%%%%%%%%%%%%%%%%%%%%%%%%%%%%%%%%%%%%%%%%%%%%%%%%

\vskip 2 cm
\begin{center}
{\large Back and Forth}
\end{center}
\noindent
\vskip .2 cm
\noindent
A particle moves back and forth with speed {\it v} on 
a straight line {\it OA} of length {\it p}, while a
second particle moves back and forth with speed {\it w}
on a straight line {\it OB} of length {\it q}. If both
particles start from {\it O} at the same time, how long 
does it take for them to cross, simultaneous and for the 
first time, a circle of radius {\it r} and center {\it O}
in the plane {\it AOB} when
\vskip .05 cm
\noindent
{\it a.} both particles move toward {\it O}?
\vskip .05 cm
\noindent
{\it b.} one particle (pick one) moves toward {\it O}, 
the other moves away from it?
\vskip .05 cm
\noindent
{\it c.} both particles move away from {\it O}?

%%%%%%%%%%%%%%%%%%%%%%%%%%%%%%%%%%%%%%%%%%%%%%%%%%%%%%%%%%%%%%%

\end{document}